\documentclass[12pt,manuscript]{aastex}
\usepackage{apjfonts}

\newcommand{\msun}{M_\odot}

\newcommand{\gapprox}{\mathrel{\mathpalette\@versim>}}
\newcommand{\lapprox}{\mathrel{\mathpalette\@versim<}}
\newcommand{\propapprox}{\mathrel{\mathpalette\@versim\propto}}

\shorttitle{Density Variations in Tycho} 
\shortauthors{WILLIAMS ET AL.}

\begin{document}

\title{Azimuthal Density Variations Around the Rim of Tycho's
  Supernova Remnant}

\author{Brian J. Williams,\altaffilmark{1}
Kazimierz J. Borkowski,\altaffilmark{2}
Parviz Ghavamian,\altaffilmark{3}
John W. Hewitt,\altaffilmark{1}
Alwin Mao,\altaffilmark{2,4}
Robert Petre,\altaffilmark{1}
Stephen P. Reynolds,\altaffilmark{2}
John M. Blondin\altaffilmark{2}
}

\altaffiltext{1}{NASA Goddard Space Flight Center, Greenbelt, MD 20771; brian.j.williams@nasa.gov}
\altaffiltext{2}{Department of Physics, North Carolina State University,
    Raleigh, NC 27695}
\altaffiltext{3}{Department of Physics, Astronomy, and Geosciences, Towson University, Towson, MD 21252}
\altaffiltext{4}{Astronomy Department, University of California, Berkeley, CA 94720}

\begin{abstract}

{\it Spitzer} images of Tycho's supernova remnant in the mid-infrared
reveal limb-brightened emission from the entire periphery of the shell
and faint filamentary structures in the interior. As with other young
remnants, this emission is produced by dust grains, warmed to $\sim
100$ K in the post-shock environment by collisions with energetic
electrons and ions. The ratio of the 70 to 24 $\mu$m fluxes is a
diagnostic of the dust temperature, which in turn is a sensitive
function of the plasma density. We find significant variations in the
70/24 flux ratio around the periphery of Tycho's forward shock,
implying order-of-magnitude variations in density. While some of these
are likely localized interactions with dense clumps of the
interstellar medium, we find an overall gradient in the ambient
density surrounding Tycho, with densities 3-10 times higher in the NE
than in the SW. This large density gradient is qualitatively
consistent with the variations in the proper motion of the shock
observed in radio and X-ray studies. Overall, the mean ISM density
around Tycho is quite low ($\sim 0.1-0.2$ cm$^{-3}$), consistent with
the lack of thermal X-ray emission observed at the forward shock. We
perform two-dimensional hydrodynamic simulations of a Type Ia SN
expanding into a density gradient in the ISM, and find that the
overall round shape of the remnant is still easily acheivable, even
for explosions into significant gradients. However, this leads to an
offset of the center of the explosion from the geometric center of the
remnant of up to $20$\%, although lower values of $10$\% are
preferred. The best match with hydrodynamical simulations is achieved
if Tycho is located at a large ($3-4$ kpc) distance in a medium with a
mean preshock density of $\sim 0.2$ cm$^{-3}$. Such preshock densities
are obtained for highly ($\ga 50$\%) porous ISM grains.


\end{abstract}

\section{Introduction}
\label{intro}

Tycho's supernova remnant (SNR; hereafter Tycho), also known as
G120.1+1.4, 3C10, and Cassiopeia B, is a member of a small subclass of
Galactic SNRs known as the ``historical supernovae'' (SNe). Tycho is
the remnant of the SN observed in 1572 \citep{stephenson02}, and was
first characterized as a ``Type I'' SN by \citet{baade45}. Light
echoes from Tycho's explosion were first discovered by \citet{rest08},
and subsequent spectroscopy of the echoes by \citet{krause08}
determined that Tycho resulted from a standard Type Ia SN. Various
distances have been proposed in the literature for Tycho; an extensive
comparison of various distance determinations is presented in
\citet{hayato10}. Many authors have adopted the distance of 2.3 kpc,
suggested by \citet{chevalier80} and \citet{albinson86}; however,
distances in excess of 4 kpc have also been reported
\citep{schwarz95}. We adopt a distance of 2.3 kpc for this work, but
also examine the effects of larger distances.

The remnant has been widely studied across the electromagnetic
spectrum. At radio wavelengths, Tycho exhibits a classic shell-type
morphology \citep{dickel82}, while at optical wavelengths, the shock
is Balmer-dominated, with emission only seen in H$\alpha$ from the
eastern and northern limbs
\citep{kamper78,ghavamian00}. \citet{ishihara10} observed the remnant
with {\it AKARI} between 9-140 $\mu$m, attributing the emission to
dust in the ISM swept up by the forward shock. They report the
possibility of as many as a few tenths of a solar mass in dust. In the
far-infrared (IR), \citet{gomez12} reported a clear detection of the
entire shell at 70 and 100 $\mu$m with {\it Herschel}, finding the
integrated flux from the remnant to be consistent with a small amount
($<10^{-2} \msun$) of warm dust at $\sim 90$ K. No clear emission
associated with Tycho was observed at 160 $\mu$m or beyond. In X-rays,
a more complicated picture emerges, with the forward shock marked by
thin rims of nonthermal synchrotron emission \citep{hwang02}. The
X-ray emission from the interior of the remnant is dominated by strong
lines of Fe, Si, and S \citep{warren05,hayato10} from the
ejecta. However, thermal X-ray emission from the forward-shocked
interstellar medium (ISM) material is conspicuously
lacking. \citet{hwang02} searched within the {\it Chandra} data for
this thermal component, only finding it in a few small sections of the
remnant, and placing upper limits elsewhere. Recently, the remnant was
detected in $\gamma$-rays, at both TeV energies with {\it VERITAS}
\citep{acciari11} and GeV energies with {\it Fermi}
\citep{giordano12}.

Tycho has generally been considered a prototypical Type Ia SN,
expanding into uniform surroundings. However, the exact density of
these surroundings has been a matter of debate. \citet{dwarkadas98}
examined Tycho in the context of one-dimensional (1D) explosion models
of a white dwarf with an exponential ejecta density profile. They
found that to approximately match the observed size and expansion
rate, an ambient density of 0.6-1.1 cm$^{-3}$ is required. A similar
result was found by \citet{badenes06}, who compare the
spatially-integrated X-ray spectrum of the western half of the remnant
with synthetic spectra produced by 1D delayed detonation hydrodynamic
models, finding a pre-shock density of $\sim$ 1 cm$^{-3}$.

On the other hand, an argument for a lower density in the ambient
medium is made by \citet{cassamchenai07}, who examine the lack of
thermal X-ray emission from the blast wave, concluding that the
azimuthally-averaged density surrounding Tycho must be $\le 0.2-0.3$
cm$^{-3}$ (although they note that systematic errors could push this
limit as high as 0.6 cm$^{-3}$). This density estimate is consistent
with that from \citet{kirshner87}, who detected broad and narrow
H$\alpha$ emission arising from ``knot g'' along the eastern limb,
finding a pre-shock density of 0.3 cm$^{-3}$. \citet{dwarkadas00}
extended the modeling of Type Ia explosions with exponential ejecta
profiles from 1D to 2D, and \citet{katsuda10} used these results, in
combination with a study of the X-ray proper motions over a baseline
of seven years, to place a limit on the density of $< 0.2$
cm$^{-3}$. Finally, \citet{lee04} suggest, based on radio observations
of a molecular cloud in the vicinity of the northeastern portion of
Tycho, that the remnant may be expanding into a density gradient
(though they do not determine numerical values for the density).

A departure from spherical symmetry is also suggested by measurements
of the proper motion of the shell. \citet{reynoso97} found azimuthal
variations of a factor of three in the proper motions at 1.4 GHz
measured around the periphery of the shell, with those in the SW being
higher than in the NE, though \citet{moffett04} reported some
modifications to these results. \citet{hughes00} confirmed these
azimuthal variations from {\it ROSAT} images, and \citet{katsuda10}
examined high-resolution {\it Chandra} images and determined that the
X-ray proper motions of the forward shock vary by about a factor of
two. Also, \citet{ghavamian00} stated that the eastern side of Tycho
must be interacting with a warm ISM cloud because optical emission
produced by photoionizing radiation from the forward shock could be
seen tracing the outer edge of the cloud. Clearly, the ambient density
surrounding Tycho is quite uncertain, and in this paper, we examine
the ISM structure as implied by broad-band flux ratios in the IR.

The {\it Spitzer Space Telescope} has provided a new window on the
Universe in the mid- and far-IR. {\it Spitzer} has returned remarkable
images of SNRs, particularly at 24 and 70 $\mu$m
\citep{blair07,hines04,temim06,williams11b}. Young SNRs, like Tycho,
are typically in the non-radiative portion of their evolution, meaning
that the temperatures in the post-shock gas are high enough that
radiative cooling via collisionally-excited lines of metals has not
yet begun to occur. In these remnants, IR emission consists of
featureless dust continuua \citep{williams12,temim12}, arising from
warm dust, heated in the post-shock gas via collisions with energetic
ions and electrons \citep{dwek87,dwek96,williams11a}. In Type Ia SNRs,
this dust is ambient ISM dust; no newly-formed ejecta dust has ever
been found in these remnants. As we will show below, the temperature
of the dust, and thus its relative emission at 70 and 24 $\mu$m, is
most sensitive to the density of the gas in the post-shock
environment. Our goal for this paper is to use the {\it Spitzer}
images of Tycho to directly measure the post-shock density at various
azimuthal locations around the shell. Nonthermal radio emission from
young SNRs does not provide a direct measurement of the density, and
thermal X-rays from the shocked ambient medium are virtually absent in
Tycho (additionally, since thermal X-ray emission scales with the
square of the density, estimates obtained from X-rays contain a
dependence on the unknown filling fraction of the gas). Optical
emission traces only the densest Balmer-dominated shocks in the NE and
N. Thus, the IR observations, and subsequent modeling shown here,
provide the best constraints yet on density variations in the ISM
surrounding Tycho. Throughout this work, we quote the azimuthal angle
as the position angle, east of north.

\section{Data}
\label{data}

\begin{figure}
\includegraphics[width=17cm]{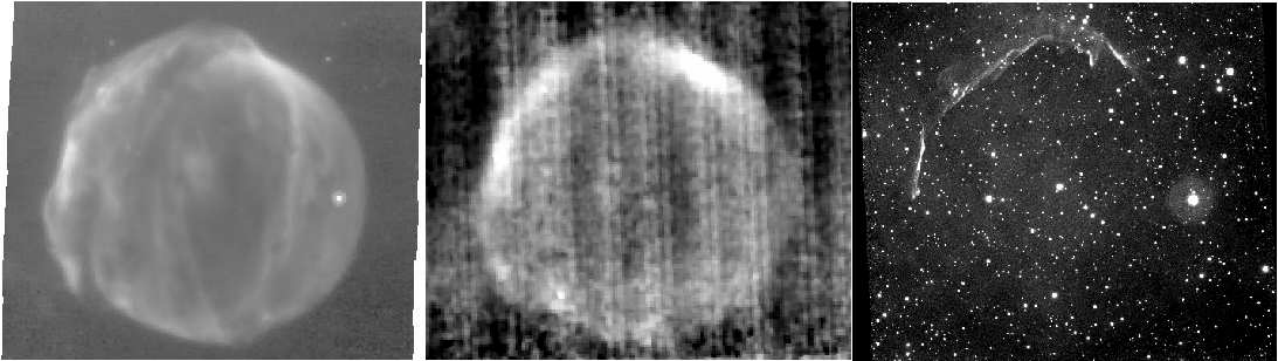}
\caption{Left: {\it Spitzer} 24 $\mu$m image. Center: {\it Spitzer} 70
  $\mu$m image. Instrument artifacts in the image introduce flux
  uncertainties at only the 8\% level (see text for details). Right:
  H$\alpha$ image from 2007, courtesy of P.F. Winkler.
\label{images}
}
\end{figure}

Tycho was observed in December of 2004 with all three instruments on
{\it Spitzer} (PI J. Rho, Program ID 3483). Here, we report only on
the 24 and 70 $\mu$m images from the Multiband Imaging Photometer for
Spitzer (MIPS). We use the Post-Basic Calibrated Data (PBCD) from the
{\it Spitzer} pipeline, version 18.12, for our analysis. We show the
24 and 70 $\mu$m images in Figure~\ref{images}. The entire
limb-brightened shell is seen in both images, and additional
filamentary structures that run roughly in a N-S direction are visible
at 24 $\mu$m. Significant variations in brightness are seen in the
remnant as well, with regions in the northwest and east having surface
brightnesses several times higher than elsewhere in the remnant. We
discuss the flux variations further in the following sections. For
further comparison, we show a high-resolution optical H$\alpha$ image
(courtesy of P.F. Winkler), which shows emission only from the eastern
and northern limbs.

\section{Results}
\label{results}

\subsection{Measurements}

\begin{figure}
\includegraphics[width=15cm]{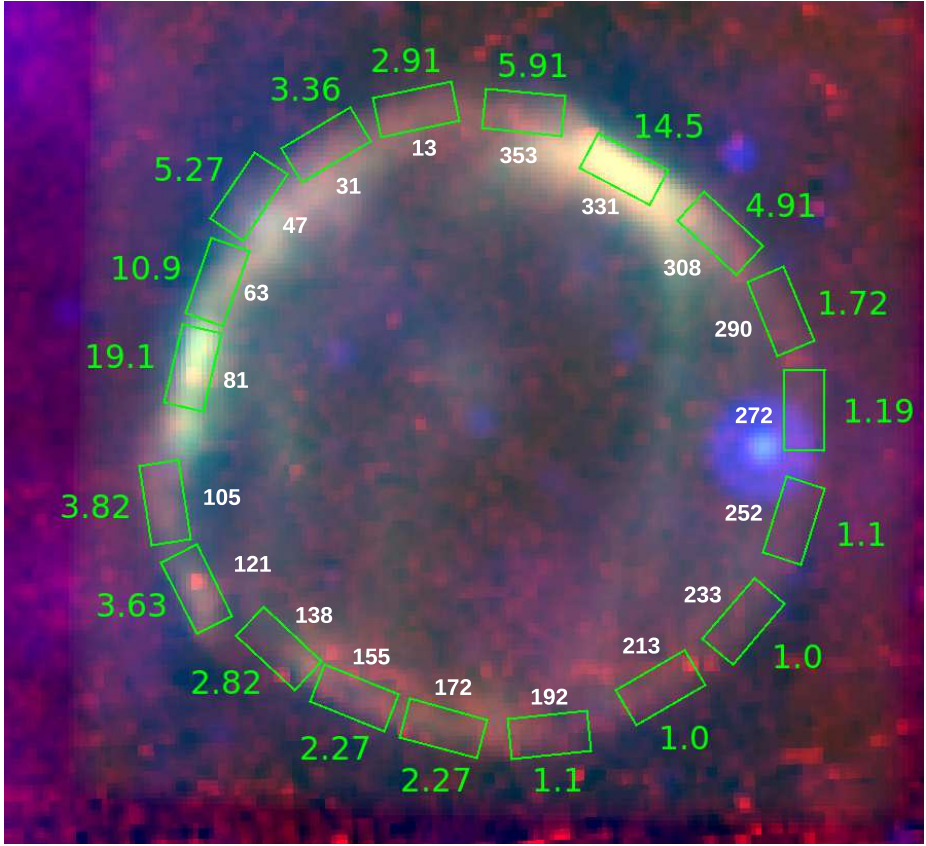}
\caption{The normalized density in various regions of Tycho, along
  with outlines of the spatial regions used for our analysis,
  superposed on a three-color IR image, with {\it Spitzer} 70 $\mu$m
  emission in red, 24 $\mu$m emission in green, and {\it WISE} 12
  $\mu$m in blue. North is up and east is to the left. The absolute
  scaling for the densities are 0.11 for the compact grain model and
  0.2 for the porous grain model (see Section~\ref{evolution} and
  Table~\ref{values} for details). Interior numbers are azimuthal
  angle, in degrees.
\label{densitymap}
}
\end{figure}

We divided the outer shell of the remnant into equally-sized segments
for our analysis, where each section is $30''$ in the radial direction
and $60''$ in the tangential direction. The radial dimension is chosen
to be slightly larger than the point-spread function of {\it Spitzer}
at 70 $\mu$m, which is approximately $20''$, to ensure that all flux
from the rim is captured within our regions. We were able to fit 19
non-overlapping regions around the periphery of the shell, ensuring
that the outer boundary of each region extends slightly beyond the
extent of the IR emission in the radial direction. We show all 19
regions, plotted on top of a 3-color IR mosaic, in
Figure~\ref{densitymap}. This mosaic contains, for visualization
purposes only, the 12 $\mu$m image from the {\it Wide-Field Infrared
  Survey Explorer (WISE)}.

To measure the IR flux ratio, we had to first decide which IR images
to use. We had six choices: 12 and 22 $\mu$m from {\it WISE}, 24 and
70 $\mu$m from {\it Spitzer}, and 70 and 100 $\mu$m from {\it
  Herschel}. We chose the {\it Spitzer} data. {\it Spitzer} images at
24 and 70 $\mu$m show clear emission from the entire shell of the
remnant. The 70 $\mu$m image suffers from regularly spaced
``striping'' artifacts related to the scan direction of the telescope
during mosaic observations. This is a well-known issue with the {\it
  Spitzer} 70 $\mu$m MIPS detector, and one that we have encountered
before \citep{sankrit10}. In that paper, we found that despite their
unsightly appearance, the stripes contribute an overall uncertainty of
only 5\% to the fluxes measured at 70 $\mu$m. Here, a similar analysis
of the variations found in ``on-stripe'' and ``off-stripe'' flux
measurements from relatively uniform regions of the remnant produced a
similar result. In all the regions we tested, most of the variations
were of order 5\%; the largest effect we found was 8\%. Thus, we did
not attempt to correct for the striping pattern; rather, we simply
added an additional 8\% uncertainty term to all measured fluxes at 70
$\mu$m. The {\it MIPS Instrument Handbook} lists calibration
uncertainties on fluxes of extended sources at 4\% and 7\% for the 24
and 70 $\mu$m detectors, respectively. With the additional 8\%
uncertainty to the 70 $\mu$m fluxes, we conservatively assume all flux
ratios to have a 16\% uncertainty.

Emission from dust grains at 12 $\mu$m arises from the very smallest
grains that undergo temperature fluctuations and which are
particularly prone to destruction through sputtering by thermal
ions. The {\it WISE} 22 $\mu$m image is essentially identical to {\it
  Spitzer} data at 24 $\mu$m, but the spatial resolution and
sensitivity of {\it WISE} are both lower than {\it Spitzer}. For this
reason, we do not use either of the {\it WISE} images in our analysis.

The {\it Herschel} data present two issues. First, while the shell of
the remnant is clearly detected at 70 $\mu$m, the detection at 100
$\mu$m is rather weak (see Figure 8 of \citet{gomez12}). Thus, our
decision is between the 70 $\mu$m data from {\it Herschel} and {\it
  Spitzer}. Ideally, we would use the {\it Herschel} data, since the
spatial resolution of {\it Herschel} is several times better than {\it
  Spitzer}, making the {\it Herschel} 70 $\mu$m image of comparable
spatial resolution to the {\it Spitzer} 24 $\mu$m image. However, the
second issue with {\it Herschel} data is calibration
uncertainty. Recent work by \citet{aniano12} has revealed significant
issues with the extended source photometry calibration of the PACS
instrument on {\it Herschel}, which contains the 70 $\mu$m
camera. Because we are only doing two-point photometry to constrain
the temperature of dust grains, it is crucial that the two fluxes that
we use are calibrated as closely to each other as possible.

With our 19 spatial regions defined, measuring the flux is
straightforward. We first convolve the 24 $\mu$m image to the
resolution of the 70 $\mu$m image using the convolution kernels
provided in \citet{gordon08}. We define a background that consists of
four separate regions, each a few arcminutes outside of the remnant,
to the NE, NW, SW, and SE of the shell. Although the background is
fairly uniform in the immediate surroundings of Tycho, we average
these four regions to create a single off-source background that we
subtract from each flux measurement (scaled to the size of our
extraction regions).

We report our flux measurements at 24 and 70 $\mu$m in
Table~\ref{measurements}. The 24 $\mu$m flux shows significant
variations as a function of azimuthal angle around the periphery of
the shell, being as much as 15 times higher in the NW than in the
SW. The 70 $\mu$m flux varies by only a factor of two around the
shell. The different behavior of the two fluxes is quite significant:
the remnant is not only brighter in some places than in others, but
{\em the ratio of the 70 $\mu$m flux to the 24 $\mu$m flux, and thus
  the temperature of the dust, varies by nearly an order of magnitude
  from one place to another in the remnant}. It is the ratio of the 70
to 24 $\mu$m flux that we fit with our models, described below, to
determine the gas density behind the shock.

\subsection{Modeling}
\label{modeling}

As previously stated, the ratio of IR fluxes resulting from emission
from warm dust grains is a diagnostic of the conditions of the X-ray
emitting plasma. We have developed spectral models for dust emission
in SNRs; we refer the reader to \citet{williams11a} for a more
complete description. Briefly, the spectrum emitted by a dust grain
immersed in a hot plasma depends on the temperature and density of
both the electrons and ions. The grain is heated by collisions with
particles, with proton and alpha particle collisions also slowly
eroding the grain via sputtering \citep{nozawa06}. These grains exist
in the ISM encountered by the forward shock wave of the SNR, and are
not newly-formed grains from the SN ejecta. Grain properties are
important as well: large grains are heated to lower temperatures than
small grains. The smallest grains (below a few nm in size) emit
radiation quickly enough that they cool back down to their ambient
temperatures before being hit again \citep{draine03}, and these
stochastic temperature fluctuations must be taken into account. Larger
grains reach an equilibrium temperature determined by the balance
between the collisional grain heating rate and the grain radiative
cooling rate. Smaller grains are more quickly destroyed via
sputtering. The optical properties of the grain are also important,
e.g., the IR spectrum from a carbonaceous grain looks different from
that of a silicate grain \citep{draine84}.

Our models take all of the above effects into account. We use the dust
grain size distributions of \citet{weingartner01}, which contain a mix
of carbonaceous and silicate grains ranging in size from 1 nm to 1
$\mu$m, in proportions appropriate for the ISM of the Milky Way. We
use their favored model for the ISM in the Milky Way, with $R_{V}$
(the ratio of visual extinction to reddening) of $3.1$ and $b_{c}$ (a
parameterization of the carbon abundance in small grains) of $6.0$,
though we note that the choice of grain-size distribution does not
significantly affect the derived fits to the IR emission
\citep{williams06}. Once the grain properties are defined, what
remains is to determine the plasma conditions. Because grains are
heated by the forward-shocked gas of roughly cosmic abundances and we
assume that the contribution of heavier elements to the gas density is
small, we can assume that the post-shock electron density is 20\%
higher than the proton density, i.e. $n_{e} = 1.2n_{p}$, while the
alpha particle density is one-tenth the proton density,
i.e. $n_{\alpha} = 0.1n_{p}$.

\subsubsection{Ion Temperature}

Determining the plasma temperatures for the various regions in Tycho
is not necessarily straightforward, but we use the following
approach. First, for the proton temperature, we assume the standard
shock jump conditions derived from the Rankine-Hugoniot conditions:

\begin{equation}
kT_{p} = \frac{3}{16}m_{p}v_{s}^{2},
\end{equation}

\noindent
where $m_{p}$ is the mass of a proton and $v_{s}$ is the shock
velocity. To determine the shock velocity, we use the proper motion
measurements from radio \citep{reynoso97} and X-ray \citep{katsuda10}
studies, scaling the expansion rate to a distance of 2.3 kpc. Both of
these studies report the proper motion of the forward shock as a
function of azimuthal angle for the entire periphery of the remnant,
and the agreement between the two is generally within 20\% (although,
in select locations, the discrepancy between the two measurements is
as high as 40\%). Because of this, we simply use an average of the
radio and X-ray values to determine the shock velocities. For the
regions in which the X-ray and radio proper motions agree to within
10\%, we assign an uncertainty of 10\% to the shock velocity,
approximately equal to the errors reported in both papers. In areas
where the discrepancy is greater than this, we use the absolute values
of the difference between the average value and that from each study
for the uncertainty. The largest uncertainty in the shock velocity
obtained in this way is 20\%. We list all shock velocities, with
uncertainties, in Table~\ref{measurements}, along with the proton
temperatures derived from these shock velocities. We assume that the
downstream energy loss of the protons to Coulomb collisions has been
minimal, consistent with the low values for temperature equilibration
found in the H$\alpha$-emitting shocks by \citet{ghavamian00}. We also
assume that alpha particle temperatures are four times higher than
proton temperatures, i.e., that ion-ion equilibration is
negligible. This is consistent with results from SN 1006, where shocks
of similar speed have been found to have minimal ion-ion equilibration
\citep{laming96}.

There are several caveats to the proton temperature
determination. First, as previously mentioned, temperature
equilibration will bring the proton and electron temperatures closer
together. However, after a fast initial rise in the electron
temperature from $\sim 10,000$ K up to a few 100,000 K, the remaining
equilibration proceeds quite slowly, and is negligible in Tycho
\citep{ghavamian01}. An additional effect is introduced by the
uncertainty in the distance to Tycho. Lastly, the Rankine-Hugoniot
conditions themselves assume that no energy is lost in the shock to
escaping particles, an assumption which breaks down in the case of
efficient cosmic-ray acceleration. Tycho may well be the site of such
acceleration, as has been proposed by \citet{warren05} and
\citet{eriksen11}. Efficient particle acceleration will lower the
post-shock proton temperature from the values we report. A factor of
two change in $kT_{p}$ solely due to cosmic-ray acceleration would
imply a compression ratio of $\sim 10$ \citep{vink10}.

However, the effects of these uncertainties in the proton temperature
on the modeling of warm dust emission are not large. Even a factor of
two variation in $kT_{p}$ only results in a 25\% change in the
inferred densities. We can quantify the dependence of the densities we
report in Section~\ref{discussion} on the uncertainty in the
distance. We assume D = 2.3 kpc, but if the distance were, for
instance, 3 kpc, the proton temperatures implied by the shock
velocities would be 70\% higher. Higher temperatures would lead to
lower densities, but the densities we report would only be lower by a
factor of 1.2. Finally, it is important to note that these
uncertainties in the proton temperature will affect only the absolute
determinations of the density, and not the relative density from place
to place in the remnant.

\subsubsection{Electron Temperature}

\citet{hwang02} report electron temperatures of around 2 keV in Tycho
based on a 49 ks {\it Chandra} observation in September of
2000. Atomic databases used in X-ray modeling codes like XSpec have
been updated significantly since 2002. We reanalyze the X-ray data for
this paper, using NEI version 2.0, augmented by custom atomic line
codes which include missing inner-shell electron transitions
\citep{badenes06}. Since Tycho was reobserved for significantly longer
in 2009, we use the new data, but repeat the analysis of
\citet{hwang02}. We also include {\it XMM-Newton} observations (PI
A. Parmar) and fit both spectra independently.

We derive a slightly lower temperature of 1.35 keV. To explore the
discrepancy between our fits and those from \citet{hwang02}, we
applied the spectral models used to fit the 2009 data to the original
2000 data, finding the same lower temperature. We obtain statistically
identical fits from the {\it XMM-Newton} data. We believe the most
likely explanation for the discrepancy between \citet{hwang02} and our
work is simply a change in the calibration of the telescope and/or
updated atomic data over the last decade. We therefore set the
electron temperature at 1.35 keV, and make the assumption that this is
the temperature in every region around the periphery of the shock.

Obviously, this assumption is unlikely to be correct everywhere along
the rim. But how much does the value of the electron temperature
matter for our dust modeling? As it turns out, it matters even less
than the proton temperature does. A factor of two difference in the
electron temperature has only a 10\% effect on the ratio of the IR
fluxes measured in the {\it Spitzer} bands. The difference between
1.35 and 2 keV, the value reported in \citet{hwang02}, has only a 5\%
effect on the density. Because the efficiency of grain heating by
electrons decreases with increasing electron temperature, values
higher than 2 keV have virtually no added effect on lowering the
densities we calculate. However, lower electron temperatures can raise
the inferred densities. The lowest observed value for the
electron/proton temperature ratio at a collisionless shock is $\sim
0.05$, found in SN 1006 \citep{vink03}. In the theoretical models of
\citet{vanadelsburg08}, shocks of greater than 1500 km s$^{-1}$ have a
minimum value of this ratio of 0.03. If we assume this value is the
minimum possible value in Tycho, this leads to a minimum electron
temperature of around 0.3 keV, which is also the minimum level of
electron heating in collisionless shocks predicted by the lower hybrid
wave model of \citet{ghavamian07}. Since electron temperatures this
low are typically not found in young SNRs, this temperature can be
regarded as a lower limit. The density inferred from a shock with
$kT_{e} = 0.3$ is a factor of 1.7 higher than that inferred from a
shock with our assumed value of 1.35 keV. Thus, while we recognize
that the assumption of a constant value of $kT_{e}$ is probably not
valid, it is unlikely that variations from this are large (given the
fact that shock speeds only vary by a factor of two in the remnant),
and in any case, even significant variations do not have a large
effect on the modeling.

\subsubsection{Density Fits}

\begin{figure}
\includegraphics[width=15cm]{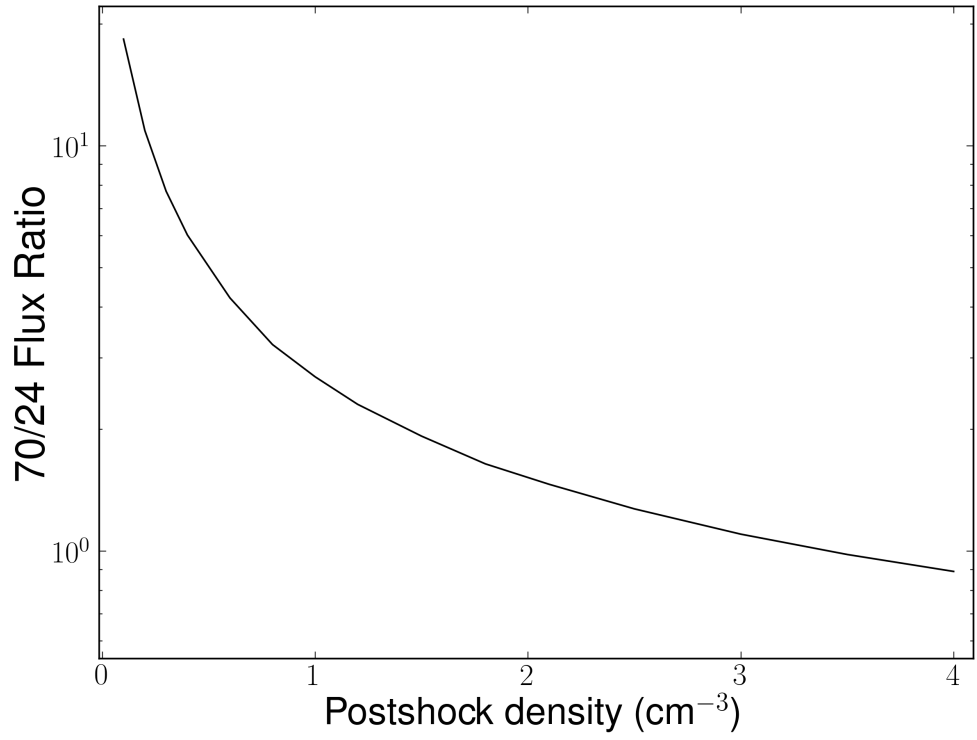}
\caption{The 70/24 $\mu$m flux ratio as a function of gas density,
  assuming electron and proton temperature held constant at 1.35 and
  20 keV, respectively. Note that this plot is on a logarithmic scale
  on the y-axis.
\label{density}
}
\end{figure}

With the electron and proton temperatures estimated for each of the 19
regions, the density of the gas in the post-shock environment is the
only remaining free parameter in our models. We adjust the density in
each region to fit the measured 70/24 flux ratio. We show in
Figure~\ref{density} the 70/24 flux ratio as a function of postshock
density, $n_{p}$, assuming constant values of $kT_{p}$ and
$kT_{e}$. The effect of the density on the flux ratio is quite large,
allowing us to determine the density within a given region with
relatively little uncertainty, for given values of the plasma
temperature. We report the values of density in Table~\ref{values} and
show them (normalized to the lowest density regions to show the
magnitude of the relative density differences in the remnant) on an
image of the remnant in Figure~\ref{densitymap}. Density uncertainties
are calculated assuming a 16\% error on the 70/24 flux ratio; see
Section~\ref{measurements}. Two inferences are immediately apparent
from the inferred densities. First, there are only three regions where
the density (reported as the post-shock proton density, $n_{p}$) is
higher than 1 cm$^{-3}$: two contiguous regions on the eastern limb,
including the ``knot g'' filament seen in the optical, and one region
in the NW. Second, aside from these three ``dense'' regions, there
appears to be an overall azimuthal gradient in the densities, with the
average in the NE being a factor of 3-5 higher than in the SW. We
discuss and interpret these results below, in
Section~\ref{discussion}.

\section{Discussion}
\label{discussion}

The densities we determine in the various regions around the forward
shock have an inverse relationship with the shock velocities measured
in these regions, as inferred from the proper motion studies. This is
expected, but is not an {\it a priori} constraint of the models. It is
also not simply a result of the effect of variations in the proton
temperature in these regions; as we showed in Section~\ref{modeling},
the effect of the proton temperature on the calculated IR flux ratio
is fairly small, and certainly not enough to account for the
variations seen in the density. Morphologically, the regions where the
highest densities are measured are also the places where H$\alpha$
emission is seen. Our models are not directly sensitive to the
pre-shock ambient density, only to that in the post-shock
environment. A standard strong shock will produce a compression ratio
of four between the pre and post-shock gas, but this should be
considered a lower limit. Efficient particle acceleration will raise
the compression ratio at the shock \citep{jones91}.

Along the eastern and northwestern portions of the remnant, the
forward shock is encountering localized denser clumps of material. In
particular, the highest density region we find in the remnant
corresponds to ``knot g'' \citep{kamper78}, the brightest optical
filament in the remnant. Our results in these few dense knots are
qualitatively consistent with the conclusions of \citet{ghavamian00}
that the shock in the regions of bright H$\alpha$ emission is
encountering denser material from a neighboring HI cloud. 

However, strong H$\alpha$ emission in Figure~\ref{images} is not
confined just to a few dense knots located at the remnant's rim, but
extends over the whole NE quadrant. The longest contiguous set of
filaments in the NE is located interior to the outermost blast
wave. Densities along these filaments are likely significantly higher
than along the remnant's rim. Because of the poorly known shock
geometry in this region of the remnant, it is not clear how the
outermost blast wave related to these optical filaments. In one
scenario, it might mark where the blast wave wraps around a
low-density periphery of a large cloud with a high density, similar to
the dense knots discussed above. Alternatively, densities along these
long optical filaments might be lower, so they would fit better into
the framework of the overall density gradient discussed below. In this
case, the magnitude of the overall density gradient would be
underestimated with the current density measurements available just at
the remnant's rim. Proper motion measurements of optical filaments are
needed to resolve these ambiguities.

\subsection{Density Gradient}

The densities we determine clearly favor a low average ISM density,
consistent with the lack of thermal X-ray emission from the blast
wave. Densities in the E and NE are 3-5 times higher than those in the
W and SW. As we will show below, hydrodynamical simulations are
broadly consistent with this, though they suggest that the magnitude
of the gradient may need to be somewhat higher, at around an order of
magnitude. The simplest explanation for this behavior is a NE-SW
density gradient in the ISM. This is consistent with the hypothesis of
\citet{lee04} that Tycho is expanding into an ISM density gradient. It
is also qualitatively consistent with studies of the X-ray emitting
ejecta by \citet{badenes06} and \citet{hayato10}, who found that the
ejecta are brighter in the E and NW than in the SW. Even if our models
for dust emission have systematic errors, the fact remains that the
temperature of the dust, as measured by differences in the IR flux
ratios, varies significantly in different locations in the remnant,
and variations in the post-shock gas density are the most plausible
way to explain this. It is unlikely that our models are more correct
on one side of the remnant than the other. Detailed multi-dimensional
modeling of the remnant's evolution and the ionization state of the
reverse-shocked ejecta is beyond the scope of this paper, but a full
understanding of Tycho will require such work.

A possible alternative explanation for the apparent density gradient
is cosmic-ray acceleration that is much more efficient on one side of
the remnant than the other. Our dust models are insensitive to this,
but to explain the density differences we see purely by an increase in
the compression ratio due to particle acceleration, the shock
compression ratio in the E and NE would have to be $\ge 15$. While
this is not beyond the realm of possibility, it does imply that $>
80$\% of the total shock energy is being put into cosmic rays
\citep{vink10}. Also, if this were the case, then the ratio of the
radius of the contact discontinuity (CD) to the forward shock (FS)
would be much higher in regions of such extreme compression
ratios. \citet{warren05} examined this ratio (CD/FS) around the
periphery of Tycho, finding evidence for particle acceleration
everywhere in the remnant (although \citet{orlando12} have suggested
that the high values of this ratio can be explained via hydrodynamic
instabilities, without need of cosmic-ray acceleration). Furthermore,
if such acceleration were taking place in regions of high density, we
would expect to see a correlation between the densities we measure and
the value of CD/FS from \citet{warren05}, yet we see no such
correlation in our data. Even when we throw out the three highest
values of density, those apparently coming from dense knots, we still
find virtually no correlation between density and CD/FS (R$^{2}$ =
0.14).

\subsubsection{Gamma-Ray Emission}

\citet{morlino12} model the $\gamma$-ray emission from Tycho by
assuming a uniform pre-shock density of 0.3 cm$^{-3}$, finding that
such a medium supports a model in which the GeV and TeV emission is
produced by the hadronic mechanism, i.e., protons accelerated at the
blast wave colliding with ambient protons to produce $\pi^{0}$
particles, which then decay into $\gamma$-rays. However,
\citet{atoyan12} find that a leptonic model of $\gamma$-rays produced
via inverse-Compton scattering of energetic electrons off of
low-energy photons fits the $\gamma$-ray data equally well. Their
model assumes a density of 0.75 cm$^{-3}$, although they note that
lower densities cannot be excluded. Only three of the nineteen regions
we model can have a pre-shock density as high as the assumed density
in either the hadronic or the leptonic models. Clearly, further study
of the $\gamma$-ray emission from Tycho is necessary, taking into
account the density structure we find here.

Interestingly, \citet{acciari11} reported a small offset in the TeV
emission from the center of the remnant, in the direction of enhanced
density to the NE. The TeV source is offset by $2.4'$ with a
statistical uncertainty of $1.4'$, while the {\it VERITAS} telescope
has a 1$\sigma$ PSF of $6'$. However, even confirmation that the
$\gamma$-ray morphology correlates with the density enhancements does
not definitively select one model of $\gamma$-ray emission, as both
leptonic and hadronic models could explain such a
morphology. $\gamma$-ray emission from bremsstrahlung and
$\pi^{0}$-decay is enhanced by the higher target density, while
inverse Compton scattering off low-energy IR or microwave photons
will also be enhanced in this region. We will fully examine the
implications of our density finding for the $\gamma$-ray emission
observed in a future publication.

\subsection{An Off-Center Explosion Site}

An additional effect of an explosion into a non-uniform ISM is that
the center of explosion will not be at the center of the resulting
remnant \citep{dohm96}. A recent analysis by \citet{kerzendorf12} of
six stars identified by \citet{ruiz04} within the center of Tycho did
not turn up any potential candidates for the companion star, thought
to exist in the single-degenerate scenario of a Type Ia SN, leading
the authors to conclude that Tycho could not be explained by such a
scenario. However, these authors searched for potential companion
stars only within a circle of radius 39$''$ centered on the center of
symmetry of the {\it Chandra} image. We report here the effects of the
density gradient that we observe on the current location of the center
of explosion with respect to the center of symmetry of the remnant. We
have performed two-dimensional (2D) hydrodynamic modeling of Type Ia
SN explosions with an exponential ejecta profile \citep{dwarkadas98}
into a density gradient, and have compared these results with analytic
solutions to the thin-shell approximation \citep{carlton11}. We report
both below. We find that the most model-independent way to infer the
offset is from the observed velocity asymmetry, i.e., the proper
motion of the shock. We assume that the observed density gradient is
in the plane of the sky.

\subsubsection{Hydrodynamic Modeling}

For our hydrodynamic modeling, we employed the same numerical methods
as described in \citet{warren13}, with the additional feature of a
density gradient in the ambient medium. The external density gradient
was of the form $\rho \propto e^{ar}$, where $r$ is a dimensionless
distance, the same as $r'$ in Equation (3), and $1/a = r_{ISM}$ is the
dimensionless ISM density scale, discussed further in the Appendix.

Specifically, we used the VH-1 hydrodynamics code to evolve the Euler
equations for an ideal gas with a ratio of specific heats of $\gamma =
5/3$ on a 2D spherical-polar grid with 300 radial zones by 900 angular
zones, providing a spatial resolution of $\Delta\theta \approx 3.5
\times 10^{-3}$ in the angular direction and slightly higher
resolution in the radial direction. The supernova ejecta were modeled
using the exponential density profile of \citet{dwarkadas98}. We
employed a moving grid to track the evolving SNR as it expanded over
five orders of magnitude, effectively removing any artifact of the
initial conditions and providing sufficient time for the
Rayleigh-Taylor instability of the contact interface between shocked
ejecta and shocked ISM to reach a quasi-steady state.

\begin{figure}
\includegraphics[width=15cm]{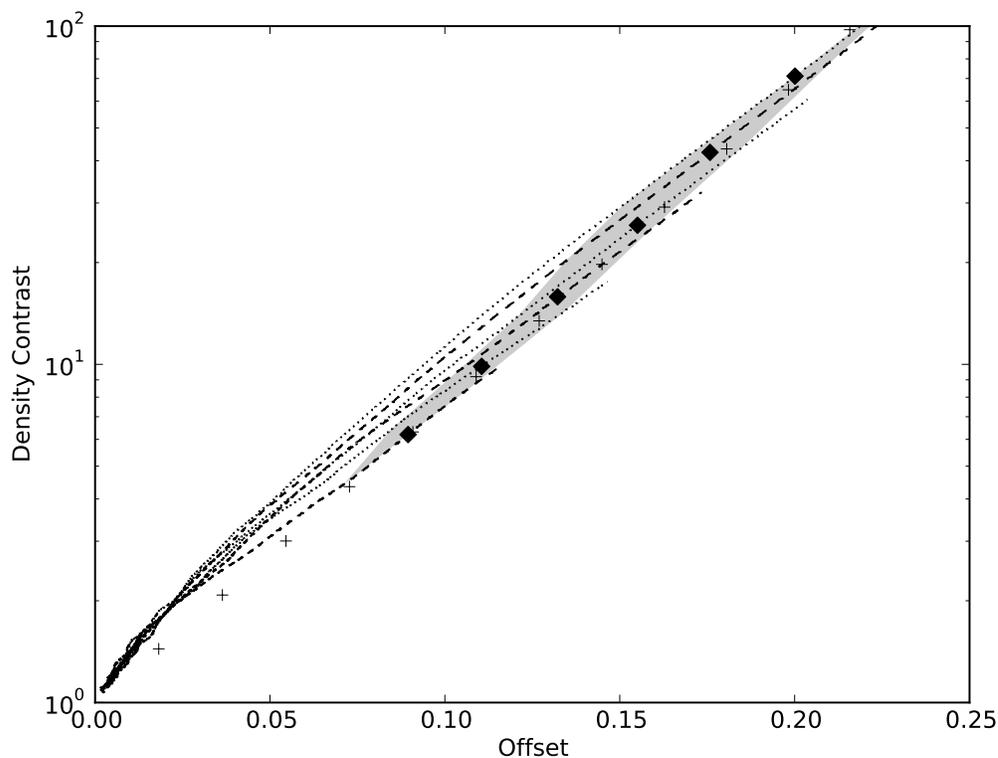}
\caption{Offset of the explosion from the geometric center of the
  remnant ($R_{max} - R_{min}$)/($R_{max}+R_{min}$) versus density
  contrast in the ISM from one side of the remnant to the other. From
  top to bottom: tracks from 2D hydrodynamical simulations with values
  of the density gradient parameter, ``$a$,'' of 1.8, 1.6, 1.4, 1.2,
  1.0, and 0.8. Diamonds show locations along each track of a
  dimensionless time of unity. The shaded gray region corresponds to
  dimensionless time between 0.7 and 1.3; Tycho should be located in
  this region. Plus signs mark analytic results for dimensionless time
  of unity, for values of ``$a$'' from 0.2 (lower-left) to 2.4
  (upper-right).
\label{contrast}
}
\end{figure}

We examined various values of $a$ to study the effects of the density
gradient on asymmetries in the shock velocities and on the remnant
shape. The values of $a$ correspond to a range of current density
contrasts at the current size of Tycho of factors of 5-100 from
maximum to minimum. An important result of our simulations is that the
ratio of the velocity semi-amplitude ((V$_{max}$ -
V$_{min}$)/(V$_{max}$ + V$_{min}$)) to the radial offset from the
center of the explosion ((R$_{max}$ - R$_{min}$)/(R$_{max}$ +
R$_{min}$)) is roughly constant at a value of about 2.2 $\pm$ 0.1 for
ages between about 300 and 700 yr. Our simulations predict, for
different values of $a$ and different ages, relations between the
radial offset and the density gradient. These relations are summarized
in Figure~\ref{contrast}, which shows that for a wide range of
gradients and ages, there is a fairly tight relation between the
density contrast across the remnant and the radial offset. One can
then use either an observed density contrast or shock proper motions
to predict the radial offset of the explosion site from the symmetry
center for a remnant.

\begin{figure}
\includegraphics[width=15cm]{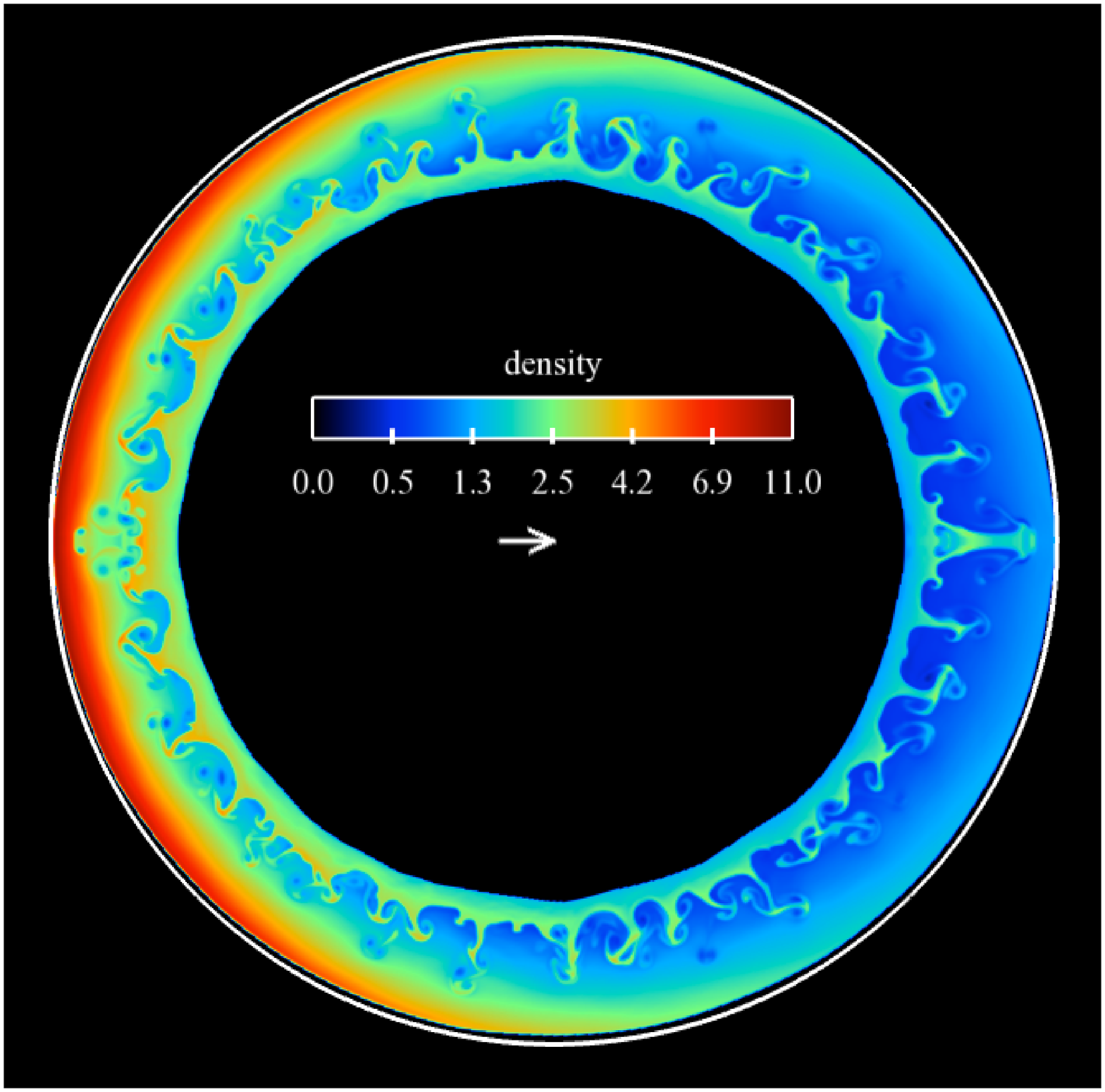}
\caption{Simulated remnant from a 2D hydrodynamical simulation at an
  age of several hundred years, expanding into an exponential density
  gradient with $a=0.95$. At the time shown, the upstream density is
  an order of magnitude higher on the left than the right. The density
  scale is in dimensionless units, and only the shocked gas is
  shown. The offset of the center of the explosion from the current
  center of symmetry of the blast wave is shown with an arrow, and
  amounts to roughly 10\%. The outer white circle is perfectly round,
  and is simply shown for comparison to illustrate the roundness of
  the simulation.
\label{hydro}
}
\end{figure}

For Tycho, Table~\ref{measurements} shows that the velocity
semi-amplitude of the averaged velocities from the radio and X-ray
measurements is 0.36 (the velocities reported in that Table assume
D=2.3 kpc, but since the velocity semi-amplitude is dimensionless, it
is also independent of distance). Dividing this by the ratio of 2.2,
given above, we obtain a radial offset of 16.5\% of the radius of the
remnant, or about $40''$ from the geometric center of the remnant to
the explosion site. The discrepancies between the radio and X-ray
measurements are relevant here. Using the X-ray data alone, the
velocity semi-amplitude is only 0.22, which leads to an offset of
$25''$, within the search radius of \citet{kerzendorf12}. Using the
radio data alone, we obtain a velocity semi-amplitude of 0.51, leading
to an offset of $1'$. Figure~\ref{hydro} shows our simulated remnant
at an age roughly corresponding to that of Tycho, and assuming a value
of $a$ of 0.95, giving a current density contrast of about an order of
magnitude. This demonstrates that the remnant (in the relatively early
evolutionary stage of Tycho) can remain remarkably round despite a
significant external density gradient. A qualitatively similar result,
for a considerably different functional form of density gradient, was
obtained by \citet{dohm96}.

The large offset inferred from the radio proper motions alone can
likely be ruled out by the results of our hydrodynamic and analytical
modeling shown in Figure~\ref{contrast}. An offset of 1$'$ would
require a density contrast of approximately 80, well above what we
measure from the IR data. Clearly, better proper motion measurements
are necessary to resolve this discrepancy. We have recently been
approved for new {\it JVLA} observations, which will expand the
baseline for proper motion measurements from the 11 years reported in
\citet{reynoso97} to $\sim 30$ years, significantly reducing the
uncertainties on the proper motions. Nonetheless, we suggest that
searches for the progenitor companion in Tycho be extended by a factor
of two beyond the area searched by \citet{kerzendorf12}.

\subsubsection{Analytic Approximation}

\citet{carlton11} present a thin-shell approximation to a blast wave
expanding into a uniform ISM, and here we extend their results to an
ambient density gradient. There are analytic solutions in the
thin-shell approximation for both linear and exponential density
gradients (see the Appendix). For the linear gradient, we derive an
approximate expression for the ratio of the dimensionless velocity
semi-amplitude to the radial offset. A full derivation is presented in
the Appendix, but the key results here are 1) the ratio from the
analytic approximation is 2.4, quite close to the 2.2 determined from
the hydrodynamic simulations above; and 2) this approximate expression
holds even for large density contrasts inferred for Tycho where use of
the exponential (instead of linear) density gradient is preferred.

\subsection{Pressure}

With the density and velocity known, we can calculate the ram pressure
in the shock, $\rho_{0} v_{s}^{2}$. We divide the post-shock number
densities by four to convert to pre-shock densities, then multiply
that by [1.4 $\times$ (1.67 $\times 10^{-24}$)] g to get a mass
density, $\rho_{0}$. We report the ram pressure for each region in
Table~\ref{values}. In the east and NW knots, the pressures are a
factor of 5-10 higher than elsewhere in the remnant, supporting the
idea of a relatively recent encounter with denser clumps, where the
pressure has not yet had time to equilibrate. Elsewhere in the
remnant, pressures in the SW are about a factor of two lower than
those in the NE, in good agreement with our hydrodynamic modeling. The
same caveat applies here about particle acceleration increasing the
compression ratio of the gas. If the compression ratio were higher in
the NW, then the ram pressures of the shock could be more equal. If
particle acceleration is constant in the remnant, only the absolute
values of the pressure will be affected, but the relative differences
will remain.

\subsection{Evolutionary State of Tycho}
\label{evolution}

The densities we report in Table~\ref{values} assume a distance to
Tycho of 2.3 kpc and ``standard'' ISM dust grains, i.e., compact,
homogeneous spheres with a mix of silicate and graphite grains in
separate populations. If we ignore the three regions coincident with
dense knots, the azimuthally-averaged value of the post-shock density
is $\sim 0.3$ cm$^{-3}$. Under the assumption of standard shock jump
conditions, this leads to an average pre-shock density, n$_{0}$, of
around 0.08-0.1 cm$^{-3}$. Since the IR emission is produced
throughout the post-shock region where densities decrease from their
value right behind the blast wave, the density ratio averaged over
that region is expected to be somewhat less than 4
\citep{williams11a}. This density, while below the upper limit of
0.2-0.3 cm$^{-3}$ reported by \citet{cassamchenai07} \&
\citet{katsuda10}, is rather low, given what is known about the
evolutionary state of Tycho.

To first-order, the average density obtained for a density gradient is
equal to the density being encountered by the forward shock in a
direction perpendicular to the gradient. For Tycho, this would
correspond to position angles of $\sim 140$ and
$320^{\circ}$. Following the scaling of hydrodynamical variables by
\citet{dwarkadas98}, we can write the dimensionless scaled radius,
$r^{\prime} \equiv r/R^{\prime}$, and the scaled time $t^{\prime}
\equiv t/T^{\prime}$ as

\begin{equation}
r^{\prime} = 0.61\ (n_{0}/0.1\ {\rm cm}^{-3})^{1/3}\ (D/2.3\ {\rm kpc})\
(M_{ej}/M_{Ch})^{-1/3},
\end{equation}

\begin{equation}
t^{\prime} = 0.82\ (n_{0}/0.1\ {\rm cm}^{-3})^{1/3}\ E_{51}^{1/2}\ (M_{ej}/M_{Ch})^{-5/6},
\end{equation}

\noindent
where M$_{ej}$ is the ejecta mass, M$_{Ch}$ is the Chandrasekhar mass,
and $E_{51}$ is the explosion energy in units of $10^{51}$
ergs. Assuming a standard Type Ia explosion where M$_{ej}$ = M$_{Ch}$,
a pre-shock density of 0.08 leads to a scaled radius of
0.57. Comparing this with the three-dimensional hydrodynamic
simulations of \citet{warren13}, a scaled radius of 0.57 for the
forward shock implies a scaled time (again, using the formalism of
Dwarkadas \& Chevalier 1998) of only $\sim 0.4$. This is too short to
explain the expansion parameter, $m$ (where r $\propto$ t$^{m}$), of
Tycho, known from proper motion measurements to be $\sim 0.53$
\citep{reynoso97,katsuda10}. Recent three-dimensional hydrodynamical
simulations of \citet{warren13} show that this can be achieved with a
scaled time of around 1, corresponding to a scaled radius also around
1-1.1. Our hydro simulations confirm this; the expansion parameter and
velocity semi-amplitude in Tycho are best matched in simulations with
a scaled time of 1.05 and a scaled radius of 1.15.

Thus, our scaled radius needs to be increased by about 70\% to match
hydrodynamic simulations of the evolutionary state of Tycho. There are
three parameters in equation (2): $n_{0}$, $D$, and $M_{ej}$. A
sub-Chandrasekhar ejecta mass would raise $r^{\prime}$, but only as
M$_{ej}^{-1/3}$, so we view this as the least viable option. If
distance alone were varied, a distance of 3.9 kpc would suffice. This
would lower the density slightly (see Section~\ref{modeling}), but
since the effect on density would be small and $r^{\prime}$ only goes
as n$_{0}^{1/3}$, the overall effect of the lowered density on
$r^{\prime}$ would be trivial. A distance of 3.9 kpc is at the high
end of the range of reported distances for Tycho (see
Section~\ref{intro}), and it is worth noting that such a distance
combined with measured proper motions would imply that shock
velocities exceed 6000 km s$^{-1}$ in some locations.

A third, perhaps most intriguing possibility, is that the densities
may be higher than we report in Table~\ref{values} due to alternative
grain models. The assumption that dust grains are compact, spherical,
solid bodies of homogeneous material makes for easier calculations,
but is unlikely to be a physically valid model \citep{shen08}. In
\citet{williams11a}, we explored the effect of porosity, or
``fluffiness,'' as well as that of a non-homogeneous grain material,
on the IR spectra produced from collisionally-heated grains. The
details of our grain model are discussed in that paper; we use an
identical model here. For grains that are 50\% vacuum, 33.5\% silicate
and 16.5\% amorphous carbon, the densities we infer from the IR flux
ratios are increased by $\sim 80\%$. This increase in the density
would raise the scaled radius to 0.7. While this is not, by itself,
enough to account for the observed evolutionary state of Tycho, it can
be combined with a more modest distance increase to only 3.3 kpc. The
resulting average density from this model would be 0.14-0.15, which is
still below the upper limits of 0.2-0.3 discussed earlier. While this
is suggestive that porous grains may, in fact, be ubiquitous in the
ISM, current knowledge of grain physics is insufficient to
conclusively rule on this issue. Grains of even higher porosity would
have a larger effect on raising the density, and grains exceeding 90\%
porosity have been suggested to fit the IR spectra of some stars
\citep{li98}. Finally, we point out that while the choice of grain
model can affect the absolute density values one determines, the
relative values are unchanged, and the inference of a density gradient
in Tycho is unaffected.

\section{Conclusions}

Observations of Tycho in the mid-IR reveal significant color
variations in the forward shock-heated interstellar dust. Since this
dust is warmed by collisions with energetic particles in the postshock
gas, the IR observations are a powerful diagnostic of the gas density
in the remnant. We analyzed IR emission as a function of azimuthal
angle around the periphery of the forward shock, finding a variation
in the ratio of the 70 $\mu$m to 24 $\mu$m flux of an order of
magnitude, implying an overall density gradient with densities higher
in the NE than in the SW by a factor of $\sim 3-5$. This ISM density
gradient is virtually independent of distance and assumed dust grain
composition. We also identify a few regions of significantly higher
density that morphologically correspond to bright H$\alpha$ emitting
regions. These are likely localized density enhancements produced
during the early stages of interaction with a nearby HI cloud.

Tycho joins the growing group of Type Ia SNRs that are not consistent
with expansion into a uniform ISM. We use two-dimensional hydrodynamic
simulations of Type Ia SN explosions into a density gradient and find
that the explosion center of the remnant must be shifted with respect
to the geometric center of the remnant by around 10\%, and possibly
higher. In Tycho, these computational simulations agree quite well
with analytic approximations. The simulations suggest that the
magnitude of the density gradient may need to be even higher than
inferred from IR observations, around a factor of 5-10, to explain the
proper motions observed in the remnant. The offset of the explosion in
the simulations is most sensitive to the observed proper motion
variations of the forward shock, so improved proper motion
measurements of Tycho are necessary for further study. Finally,
simulations show that remnants expanding into a significant density
gradient can remain remarkably round.

The mean upstream densities in Tycho are relatively low ($\sim
0.1-0.2$ cm$^{-3}$, depending on the assumed dust model), consistent
with the lack of thermal X-ray emission from most places along the
forward shock in Tycho (with the exception of the NW). The densities
inferred from models of standard ISM dust grains are too low to
explain the evolutionary state of Tycho. Porous grains provide a
potential resolution to this, as inferred densities from such grain
models are significantly higher. A distance of around 3.5 kpc would
also resolve this issue. Our results suggest that Tycho is far from a
spherically symmetric, homogenous remnant, and multi-dimensional
modeling is required for a fuller understanding.

{\it Facilities:} \facility{Chandra}, \facility{Spitzer},
\facility{XMM-Newton}

\acknowledgements

We thank the anonymous referee for a careful reading of the text. We
thank P.F. Winkler for the H$\alpha$ image. This work is partly based
on observations made with the Spitzer Space Telescope, which is
operated by the Jet Propulsion Laboratory, California Institute of
Technology under a contract with NASA. This work was supported by NASA
Archival Data Analysis Program grant
12-ADAP12-0168. B.J.W. acknowledges support from the NASA Postdoctoral
Program Fellowship. K.J.B. and S.P.R. acknowledge support from NASA
through grant NNX11AB14G. A.M. acknowledges support through NSF's REU
award AST-1032736 to NC State University.

\newpage
\clearpage

\begin{deluxetable}{lccccc}
\tablecolumns{6} 
\tablewidth{0pc} 
\tabletypesize{\footnotesize}
\tablecaption{Measurements} \tablehead{ \colhead{Deg.} & 24 $\mu$m
  Flux & 70 $\mu$m Flux & 70/24 Flux Ratio & $V_{s}$ (km s$^{-1}$) 
& T$_{\rm P}$ (keV)}

\startdata

13 & 201 & 1319 & 6.96 $\pm 1.1$ & 3660 $\pm 366$ & 26.0\\
31 & 223 & 1373 & 5.97 $\pm 0.93$ & 3310 $\pm 331$ & 21.3\\
47 & 226 & 1358 & 5.42 $\pm 0.84$ & 2360 $\pm 236$ & 10.9\\
63 & 514 & 1592 & 3.50 $\pm 0.54$ & 1920 $\pm 378$ & 7.1\\
81 & 1093 & 1739 & 2.09 $\pm 0.32$ & 2210 $\pm 378$ & 9.5\\
105 & 279 & 1394 & 6.07 $\pm 0.94$ & 3110 $\pm 311$ & 18.8\\
121 & 315 & 1441 & 5.99 $\pm 0.93$ & 3510 $\pm 351$ & 24.0\\
138 & 198 & 1319 & 7.65 $\pm 1.2$ & 3330 $\pm 333$ & 21.6\\
155 & 181 & 1361 & 8.79 $\pm 1.4$ & 3480 $\pm 538$ & 23.6\\
172 & 184 & 1369 & 9.21 $\pm 1.4$ & 3240 $\pm 486$ & 20.4\\
192 & 103 & 1281 & 15.10 $\pm 2.3$ & 3780 $\pm 378$ & 27.8\\
213 & 99 & 1304 & 15.70 $\pm 2.4$ & 4060 $\pm 634$ & 32.1\\
233 & 94 & 1339 & 15.92 $\pm 2.5$ & 3980 $\pm 418$ & 30.9\\
252 & 88 & 1211 & 14.33 $\pm 2.2$ & 3920 $\pm 648$ & 29.8\\
272 & 96 & 1248 & 13.70 $\pm 2.1$ & 3850 $\pm 385$ & 28.8\\
290 & 110 & 1168 & 10.58 $\pm 1.6$ & 3700 $\pm 370$ & 26.6\\
308 & 346 & 1499 & 4.34 $\pm 0.67$ & 3580 $\pm 358$ & 24.9\\
331 & 1362 & 2113 & 1.98 $\pm 0.31$ & 3200 $\pm 320$ & 19.9\\
353 & 273 & 1426 & 5.02 $\pm 0.78$ & 2380 $\pm 238$ & 11.0\\

\enddata

\tablecomments{Deg. = Azimuthal angle, east of north. All fluxes
  reported in milliJanskys. Errors on fluxes are 4\% for 24 $\mu$m and
  15\% for 70 $\mu$m (see Section~\ref{measurements} for
  details). $V_{s}$ = shock velocity, averaged from radio and X-ray
  measurements and assuming D=2.3 kpc. T$_{\rm P}$ = proton
  temperature.}
\label{measurements}
\end{deluxetable}

\newpage

\begin{deluxetable}{lcc}
\tablecolumns{3}
\tablewidth{0pc}
\tabletypesize{\footnotesize}
\tablecaption{Results}
\tablehead{
\colhead{Deg.} & Density (cm$^{-3}$) & Pressure (dyne cm$^{-2}$)}

\startdata

13 & 0.32$^{0.40}_{0.27}$ & 2.51$^{3.16}_{2.01}$ $\times 10^{-8}$\\
32 & 0.37$^{0.46}_{0.31}$ & 2.37$^{2.99}_{1.90}$ $\times 10^{-8}$\\
47 & 0.58$^{0.72}_{0.48}$ & 1.89$^{2.38}_{1.52}$ $\times 10^{-8}$\\
63 & 1.2$^{1.5}_{1.0}$ & 2.59$^{3.39}_{1.92}$ $\times 10^{-8}$\\
81 & 2.1$^{2.6}_{1.7}$ & 6.00$^{7.77}_{4.55}$ $\times 10^{-8}$\\
105 & 0.42$^{0.52}_{0.35}$ & 2.37$^{2.99}_{1.90}$ $\times 10^{-8}$\\
121 & 0.4$^{0.50}_{0.33}$ & 2.88$^{3.62}_{2.31}$ $\times 10^{-8}$\\
138 & 0.31$^{0.38}_{0.26}$ & 2.00$^{2.52}_{1.61}$ $\times 10^{-8}$\\
155 & 0.25$^{0.31}_{0.21}$ & 1.77$^{2.28}_{1.36}$ $\times 10^{-8}$\\
172 & 0.25$^{0.31}_{0.21}$ & 1.53$^{1.97}_{1.18}$ $\times 10^{-8}$\\
192 & 0.12$^{0.15}_{0.10}$ & 1.00$^{1.26}_{0.80}$ $\times 10^{-8}$\\
213 & 0.11$^{0.14}_{0.09}$ & 1.06$^{1.36}_{0.81}$ $\times 10^{-8}$\\
233 & 0.11$^{0.14}_{0.09}$ & 1.02$^{1.29}_{0.82}$ $\times 10^{-8}$\\
252 & 0.12$^{0.15}_{0.10}$ & 1.08$^{1.39}_{0.82}$ $\times 10^{-8}$\\
272 & 0.13$^{0.16}_{0.11}$ & 1.13$^{1.42}_{0.91}$ $\times 10^{-8}$\\
290 & 0.19$^{0.24}_{0.16}$ & 1.52$^{1.92}_{1.22}$ $\times 10^{-8}$\\
308 & 0.54$^{0.67}_{0.45}$ & 4.05$^{5.10}_{3.25}$ $\times 10^{-8}$\\
331 & 1.6$^{2.0}_{1.3}$ & 9.58$^{12.1}_{7.69}$ $\times 10^{-8}$\\
353 & 0.65$^{0.81}_{0.54}$ & 2.15$^{2.71}_{1.73}$ $\times 10^{-8}$\\

\enddata

\tablecomments{Densities are postshock. Pressure calculation assumes
  compression ratio of 4 at the shock front. Both assume standard ISM
  dust grain models of \citet{weingartner01} and D=2.3 kpc (see text
  for details of dependence on these quantities).}
\label{values}
\end{deluxetable}

\appendix
\section{Thin-Shell Solution}

For young Type Ia SNRs, 3-D hydrodynamical simulations by 
\citet{warren13} showed that the average radius  
of the contact interface between the shocked ejecta and the shocked ambient 
medium depends only weakly on details of the postshock flow. It then becomes
possible to use a thin-shell approximation to follow expansion of the remnant.
In this approximation, shocked ejecta and shocked ambient medium are 
assumed to reside in an infinitely-thin shell located at radius $R_s$ at 
time $t$ after the explosion and moving with 
velocity $V_s$. 
The velocity $v$ of the freely-expanding ejecta at the reverse 
shock is equal to $R_s/t$. In the exponential ejecta model of Dwarkadas 
\& Chevalier (1998), the shocked ejecta mass $M_{ej}$ is equal to
\begin{equation}
\frac {M_{ej}}{M_{ej}^{tot}} = \frac {1}{2} \left[\left(\frac {v}{v_e} \right)^2 + 
2 \frac {v}{v_e} + 2 \right] \exp \left( -\frac {v}{v_e} \right),
\label{ejectamass}
\end{equation}
where $M_{ej}^{tot}$ is the total ejecta mass and $v_e$ is the 
exponential velocity scale ($\rho_{ej} \propto \exp (-v/v_e)$). 
The shell momentum $P_s$ is equal to the momentum of the shocked ejecta
\begin{equation}
\frac {P_s}{vM_{ej}} = \frac {\left( v/v_e \right)^2}{\left( v/v_e \right)^2 +
2 v/v_e + 2} + \frac {3v_e}{v},
\label{shellmomentum}
\end{equation}
since the ambient ISM is assumed to be at rest. The shell velocity $V_s$ is 
equal to the shell momentum divided by the total shell mass, 
$P_s/(M_{ej}+M_{ISM})$ ($M_{ISM}$ is the shocked ISM mass). Equation of motion 
for the shell is $dR_s/dt = V_s$, which can be transformed to
\begin{equation}
\frac {v}{R_s} \frac {d R_s}{d v} = \left( 1 - 
\frac {vM_{ej} \left( 1 + M_{ISM}/M_{ej} \right)} {P_s} \right)^{-1}
\label{shellequation}
\end{equation}
by changing the independent variable from $t$ to $v$. For uniform ISM 
($M_{ISM} \propto R^3$), \citet{carlton11} reported an exact solution to 
equations~(\ref{ejectamass}--\ref{shellequation}):
\begin{equation}
r_0 = \left[1+4 \frac {v_e}{v}+6\left( \frac{v_e}{v} \right)^2 \right]^{1/3} 
\left( \frac{2v}{v_e} \right)^{1/3} \exp \left(-\frac{v}{3v_e} \right),
\label{radiusuniform}
\end{equation}
where the dimensionless radius $r_0$ is now normalized to unity for
the swept-up ISM mass equal to $M_{ej}^{tot}$, as in
\citet{dwarkadas98}. The dimensionless time is equal to
$r_0\left(v/3^{1/2}2v_e\right)^{-1}$ (also in \citet{dwarkadas98}
units).

Equations~(\ref{ejectamass}--\ref{shellequation}) can be solved analytically 
for more complex ISM density distributions than simple power laws in $R$ 
discussed by \citet{carlton11}. We consider here an exponential density 
distribution $\rho_{ISM} \propto \exp (ar)$, where $1/a = r_{ISM}$ is the 
dimensionless ISM density scale, negative (positive) for density decreasing 
(increasing) with radius. For $|ar| << 1$, the exponential density 
profile becomes linear, so solutions discussed below encompass both linear and 
exponential profiles. The swept-up mass is
\begin{equation}
m_{ISM} = 3 r^3 \left[ \left( \frac {1}{ar} - \frac {2}{\left( ar \right)^2} 
+ \frac {2}{\left( ar \right)^3} \right)  e^{ar} - 
\frac {2}{\left( ar \right)^3} \right]
\label{massism}
\end{equation}
in dimensionless units. By changing variables from $v$ and $R$ to 
$y=(ar_0)^3$ and $x=ar$ in equations~(\ref{ejectamass}--\ref{shellequation}),
we arrive at
\begin{equation}
x \frac {d y}{d x} = 4a^3m_{ISM} - y = 12 \left[ \left( x^2 - 2x + 
2 \right) e^x - 2 \right] - y.
\label{shellequationexponential}
\end{equation}
A solution
\begin{equation}
y = \frac {12}{x} \left[ \left( x^2 - 4x + 6 \right) e^x - 
6 - 2x \right]
\label{shellsolutionexponential}
\end{equation}
relates the scaled shell radius $x=ar$ to the scaled swept-up ISM mass 
$y=(ar_0)^3$ for ejecta entering the reverse shock with the same 
free-expansion velocity $v$ for both exponentially- and uniformly-distributed 
ambient medium. Expansion in Taylor series at $x=0$ gives
\begin{equation}
y = x^3 + \frac {3x^4}{5} + \frac{x^5}{5} + \ldots,
\label{shellsolutionlinear}
\end{equation}
where the first two terms on the right-hand side provide an exact solution 
for the linear density gradient.

Displacement $\delta r$ between solutions (\ref{shellsolutionexponential}) and 
(\ref{radiusuniform}), evaluated at the same time $t$ but with different 
ejecta velocities $v$ and $v_0$, respectively, is given by equations
\begin{eqnarray}
\frac {\delta r}{r_0} = \frac {r(v)-r_0(v_0)}{r_0(v_0)} = \frac {v_0-v}{v}, &
v_0 r(v) = v r_0(v_0).
\label{equationsnonlinear}
\end{eqnarray}
Generally, there is no simple explicit solution for $v_0$ and $\delta r$, 
but for small displacements a Taylor series expansion gives
\begin{equation}
\frac {\delta r}{r_0} = - \frac {3z\left( 1+4z+6z^2 \right) a} 
{5 \left[ 1+6z \left( 1+3z+4z^2 \right) \right]},
\label{displacement}
\end{equation}
where $z \equiv v_e/v_0$. The dimensionless offset 
$(R_{max}-R_{min})/(R_{max}+R_{min})$ between the center of the remnant and the 
true explosion center is then equal to $|\delta r/r_0|$. The dimensionless 
velocity semiamplitude $(V_{max}-V_{min})/(V_{max}+V_{min})$ is also linear in $a$, 
so their ratio 
\begin{equation}
\frac {\left( V_{max} - V_{min} \right) / \left( V_{max} + V_{min} \right)}
{\left( R_{max} - R_{min} \right) / \left( R_{max} + R_{min} \right)} = 
2 + \frac {3z\left( 1+8z+24z^2+24z^3+12z^4 \right)}
{\left[ 1+6z \left( 1+3z+4z^2 \right) \right] 
\left[ 1+3z \left( 1+2z+2z^2 \right) \right]}
\label{velocitytoradius}
\end{equation}
is independent of the magnitude of the density gradient. For young SNRs such 
as Tycho, the second term on the right-hand side of this equation varies only 
slowly with $z$, so at the age of Tycho the ratio between the dimensionless 
velocity semiamplitude and the dimensionless offset may be considered 
independent of age and equal to 2.4.

\end{document}